\documentclass[aip,reprint,amsmath,amssymb]{revtex4-1}
\usepackage{color}
\usepackage{graphicx}
\usepackage[]{enumerate}
\usepackage{bm}
\usepackage{subfig}

\DeclareBoldMathCommand{\bV}{V}
\DeclareBoldMathCommand{\bv}{v}
\DeclareBoldMathCommand{\bx}{x}
\DeclareBoldMathCommand{\by}{y}
\DeclareBoldMathCommand{\bz}{z}
\DeclareBoldMathCommand{\br}{r}
\DeclareBoldMathCommand{\bb}{b}
\DeclareBoldMathCommand{\be}{e}
\DeclareBoldMathCommand{\bB}{B}
\DeclareBoldMathCommand{\bE}{E}
\DeclareBoldMathCommand{\bk}{k}
\DeclareBoldMathCommand{\bA}{A}
\DeclareBoldMathCommand{\bJ}{J}

\newcommand\Alfven{Alfv\'en }
\newcommand\Alfvenic{Alfv\'enic }

\begin{document}
\title{Evidence of Critical Balance in Kinetic \Alfven Wave  Turbulence Simulations}
\author{J. M. TenBarge}
\email{jason-tenbarge@uiowa.edu}
\author{G. G. Howes}
\affiliation{Department of Physics and Astronomy, University of Iowa, Iowa City, IA, USA}
\date{\today}
\begin{abstract}
A numerical simulation of kinetic plasma turbulence is performed to assess the applicability of critical balance to kinetic, dissipation scale turbulence. The analysis is performed in the frequency domain to obviate complications inherent in performing a local analysis of turbulence. A theoretical model of dissipation scale critical balance is constructed and compared to simulation results, and excellent agreement is found. This result constitutes the first evidence of critical balance in a kinetic turbulence simulation and provides evidence of an anisotropic turbulence cascade extending into the dissipation range. We also perform an Eulerian frequency analysis of the simulation data and compare it to the results of a previous study of magnetohydrodynamic turbulence simulations.
\end{abstract}

\maketitle
\section{Introduction}

Plasma turbulence is ubiquitous in a wide range of space and
astrophysical environments, playing a fundamental role in transferring
energy from the large scales at which the turbulence is driven to the
small scales at which the turbulence is dissipated. Developing a
detailed understanding of plasma turbulence is one of the key goals of
the space physics and astrophysics communities.

One of the central tenets of plasma turbulence is the concept of
critical balance. Critical balance is the supposition that the time
scale associated with linear fluctuations of Alfv\'{e}n waves, $\omega
= k_\parallel v_A$, is of the same order of magnitude as time scale
associated with the nonlinear cascade of energy, $\omega_{nl} \simeq
k_\perp v_\perp$, where $v_{A}$ is the \Alfven speed, $v_\perp$ is
the perpendicular fluctuation velocity, and parallel and perpendicular
are defined with respect to the direction of the local mean magnetic
field \cite{Goldreich:1995,Goldreich:1997,Maron:2001}.  Critical
balance leads to the prediction of an anisotropic cascade of energy in
wavevector space, where magnetic energy cascades at different rates
parallel and perpendicular to the local mean magnetic field.

Although the original predictions of critical balance pertain only to
a cascade of Alfv\'{e}n waves in the magnetohydrodynamic (MHD) limit
of the inertial range, the theory can be extended to scales smaller
than the ion gyroradius, at which wave-particle damping and collisions
become important
\cite{Cho:2004,Galtier:2006,Howes:2008b,Schekochihin:2009}. The latter region
is often referred to as the \emph{dissipation range} of plasma
turbulence, where it is proposed that the cascade of Alfv\'{e}n waves
transitions to a cascade of kinetic Alfv\'{e}n waves (KAW)
\cite{Howes:2008a,Howes:2008b,Schekochihin:2009}.

The anisotropic scaling of the magnetic field energy has been observed
in the inertial range portion of the solar wind through the use of
wavelets or second-order structure functions to discern the local mean
magnetic field direction, e.g.,
\cite{Horbury:2008,Podesta:2009a,Chen:2010,Wicks:2010a,Luo:2010}.
\citet{Horbury:2008} demonstrated that critical balance fits the solar
wind observations well in the directions parallel and perpendicular to
the local mean magnetic field, and \citet{Forman:2011} went further,
demonstrating that solar wind observations follow the predictions of
critical balance for all angles between parallel and perpendicular.

Critical balance and its physical consequences have also been tested
and verified in many numerical turbulence simulations. It has been
tested extensively in MHD simulations, e.g.,
\cite{Cho:2000,Maron:2001,Cho:2002,Cho:2003,Oughton:2004}, and at
smaller, dissipation range scales in electron MHD (EMHD) simulations,
e.g., \cite{Cho:2004,Cho:2009}. However, all of the previous studies
have been performed with fluid codes that cannot capture wave-particle
interactions nor accurately model collisional effects, both of which
play important roles in dissipating turbulence at scales below the ion
gyroradius.

To evaluate whether critical balance persists in the dissipation
range, we consider here a detailed study of a numerical simulation of
dissipation scale turbulence performed using AstroGK, the
Astrophysical Gyrokinetics Code, developed to study kinetic turbulence
in astrophysical environments. Rather than examining the energy
distribution in wavenumber space, we  perform our analysis in the
frequency domain to obviate some of the difficulties inherent in
performing a local analysis of turbulence, which are discussed in \S
\ref{sec:disc}. The frequency is used as a proxy for the 
parallel wavenumber to determine whether or not an anisotropic
cascade, consistent with that predicted for critically balanced
kinetic \Alfven wave turbulence, exists in the dissipation range. An
Eulerian frequency analysis of the AstroGK simulation will also be
compared to a similar study by \citet{Dmitruk:2009} performed using a
MHD simulation.

\section{Simulation Parameters}\label{sec:sim}

A detailed description of AstroGK and the results of linear and
nonlinear benchmarks are presented in \citet{Numata:2010}, so we 
provide here only a brief overview.

AstroGK is an Eulerian slab code with triply periodic boundary
conditions that solves the electromagnetic gyroaveraged Vlasov-Maxwell
five dimensional system of equations. It solves the gyrokinetic
equation and gyroaveraged Maxwell's equations for the perturbed
gyroaveraged distribution function, $h_s(x,y,z,\lambda,\epsilon)$, for
each species $s$ (protons and electrons), the parallel vector
potential, $A_z$, and perturbed parallel magnetic field, $\delta B_z$,
and the scalar potential, $\phi$ \cite{Frieman:1982,Howes:2006}. The
simulation domain is elongated along the direction of the equilibrium
magnetic field $\mathbf{B}_0=B_0\hat{\bz}$.  Velocity space
coordinates are related to the energy, $\epsilon = v^2/2$, and pitch
angle, $\lambda = v_\perp^2 / v^2$. The equilibrium distribution for
both species is treated as Maxwellian, and a realistic mass ratio,
$m_p / m_e = 1836$, is used. The $x-y$~plane is treated
pseduospectrally, and an upwinded finite-differencing approach is
employed for the $z$-direction. Velocity space is evaluated following
Gaussian quadrature rules. Linear terms are evolved implicitly in
time, while nonlinear terms are evolved explicitly by a third-order
Adams-Bashforth method. Collisions are treated using a fully
conservative, linearized, and gyroaveraged collision operator
\cite{Abel:2008,Barnes:2009}.

To represent average solar wind conditions observed at $\simeq 1$~AU,
we choose plasma parameters $\beta_{i} = 1$ and $T_{i} = T_{e}$, where
$\beta_{i} = v_{ti}^{2} / v_{A}^{2}$, $v_A = B_{0} / \sqrt{4\pi n_{0i}
m_{i}}$, and $v_{ti} = \sqrt{2 T_{i} / m_{i}}$ is the ion thermal
speed. We output the electromagnetic field information at fixed time
intervals. Since this diagnostic is data intensive, we choose to
perform the simulation on a numerical grid smaller than the largest
simulations performed with AstroGK
\cite{Howes:2011b}. Specifically, we use a numerical grid of $(n_{x},
n_{y}, n_{z}, n_{\epsilon}, n_{\lambda}, n_{s}) =$
$(32,32,64,16,16,2)$, where $n_{\epsilon}$, $n_{\lambda}$, and $n_{s}$
are the number of grid points in energy, pitch angle, and species
respectively. The spatial extent of the domain is $(L_{x}, L_{y},
L_{z}) = 2 \pi (\rho_{i}, \rho_{i}, a_{0})$, where $\rho_{i} = v_{ti}
/ \Omega_{i}$ is the ion gyroradius, $\Omega_i = e B_0 / m_i c $ is
the ion gyrofrequency, and $a_{0}$ determines the domain elongation
and is chosen by assuming a critically balanced Alfv\'enic inertial
range: $\delta = \rho_{i} / a_{0} = ({k}_{i} \rho_i)^{1/3} ({k}_{\perp
0}\rho_i)^{2/3}$, where $k_{i}$ is the wavenumber corresponding to the
outer scale of the turbulent cascade (much larger than our simulation
domain) at which the turbulent system we are modelling is physically
driven, and $k_{\perp0}$ is the perpendicular wavenumber corresponding
to the simulation domain size, the largest resolved length scale in
the simulation (sub-script $0$ is used throughout to indicate
domain-scale quantities). Based on in situ measurements, we choose ${k}_{i}
\rho_i = 10^{-4}$ for the solar wind at $1$~AU
\cite{Howes:2008b}. Assuming this value for $k_{i}$ implies $\delta
\simeq 1/20$, yielding a simulation domain with the $z-$direction
elongated by a factor of $1 / \delta=20$.

Normalization of the time $\hat{t}=t/\tau_t$ and frequency
$\hat{\omega}=\omega \tau_t$ uses the parallel thermal time, $\tau_{t}
= a_{0}/v_{ti} = a_{0} / (v_{A} \sqrt{\beta_{i}})$, where $\beta_{i} =
1$.  The normalized linear kinetic \Alfven wave frequency at the
largest scale of our simulation is $\hat{\omega}_0 = 1.1366$,
determined from a linear gyrokinetic dispersion relation solver
\cite{Howes:2006}. The corresponding normalized domain-scale time is given by 
\begin{equation}
\hat{\tau}_{0} = \frac{2\pi}{\hat{\omega}_{0}} \simeq 5.53,
\end{equation} 
Electromagnetic field data is output every $\Delta t=0.02 \tau_{t}$,
resulting in a Nyquist frequency of $\hat{\omega}_{Nq} \sim 160$.

We drive our simulation with an oscillating Langevin antenna driving
the parallel vector potential. Details regarding the driving antenna
are provided in TenBarge et al. (2011) \cite{TenBarge:2011b}, so here
we specify only the antenna parameters. We drive the largest four
independent modes of our simulation domain, $(k_{x}, {k}_{y}, {k}_{z}
/ \delta) \rho_i = (\pm1,0,\pm1)$ and $(0,\pm1,\pm1)$, with a
frequency $\omega_{a} \simeq \omega_{0}$ and decorrelation rate
$\gamma_{a} \simeq 0.7 \omega_{0}$. The antenna amplitude, $A_{\parallel 0}$, is
chosen to satisfy critical balance at the domain-scale,
\begin{equation}
A_{\parallel 0} = \frac{\omega_{0} \sqrt{4 \pi n_{0i} m_{i}}}{k_{\perp}^{2} C_{2} 
\overline{\omega}},
\end{equation}
where $C_{2} \simeq 1$ is a Kolmogorov constant and an analytical
estimate of the linear \Alfven/kinetic \Alfven wave frequency normalized
to $k_\parallel v_A$ is given by
\begin{equation}\label{eq:disprel}
\overline{\omega} = \frac{\omega}{k_\parallel v_A} = \sqrt{1+\frac{(k_\perp \rho_i)^2}{\beta_i +  2/(1 + T_e/T_i)}}.
\end{equation}

\begin{figure}[top]
		\includegraphics[width=\linewidth]{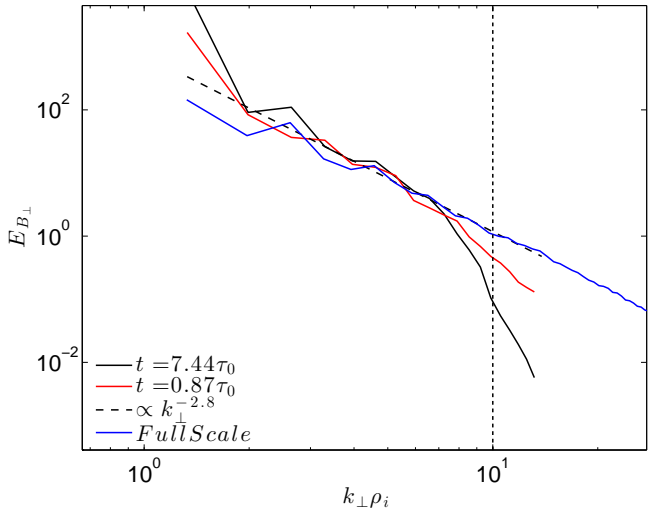}
      \caption{(Color online) Perpendicular magnetic energy spectra at the beginning (red) and end (solid black) of the frequency analysis compared to a larger, higher resolution simulation with similar plasma parameters (blue). The dashed line corresponds to a spectral slope of $-2.8$, and the vertical dotted line corresponds to the maximum fully resolved perpendicular scale.}\label{fig:perp_spec2}
\end{figure}

Plots of the perpendicular magnetic energy spectrum at the beginning
(red) and end (solid black) of the analysis included herein can be
found in Figure \ref{fig:perp_spec2}. The analysis begins after the
cascade has become well-developed, which corresponds to $t = 0.87
\tau_0$, and the analysis ends at $t = 7.44 \tau_0$. The vertical
dotted line at $k_\perp \rho_i = 10$ corresponds to the maximum fully
resolved perpendicular mode of the simulation. Beyond that point,
hypercollisionality acts to remove energy from the system and steepens
the spectrum. Due to the hypercollisionality, the domain of validity
is limited to $k_\perp \rho_i \lesssim 8$. The saturated spectra have
a spectral index $\sim -2.8$. This spectral index agrees with larger
scale numerical simulations with the same plasma parameters
\cite{Howes:2011b}. This demonstrates that, although the dynamic range
of the simulation under consideration is rather limited, the overall
behaviour is consistent with a larger scale simulation.

\section{Critical Balance Theory}\label{sec:cbt}

The strength of turbulence can be characterized by the ratio of the
nonlinear cascade rate or perpendicular eddy turn-around time,
$\omega_{nl} \simeq k_{\perp} v_{\perp}$, to the linear frequency,
\begin{equation}
\chi = \frac{\omega_{nl}}{\omega} = \frac{k_{\perp} v_{\perp}}{k_{\parallel} v_A}.
\end{equation}
When $\chi \ll 1$, the turbulent fluctuations exist for several
turn-around times at a given scale before their energy is cascaded to
smaller scales---many wave-wave ``collisions'' are required to cascade
their energy \cite{Iroshnikov:1963,Kraichnan:1965}. 
This situation, known as weak turbulence \cite{Sridhar:1994}, generates a
cascade of energy in only the perpendicular direction
\cite{Montgomery:1981,Shebalin:1983}. Therefore, $\chi$ grows 
with increasing perpendicular wavenumber, strengthening the nonlinear
interactions. Once $\chi \simeq 1$, the timescales of the
nonlinear and linear processes become equal, and the turbulence is
said to be critically balanced
\cite{Goldreich:1995,Goldreich:1997}.

It is interesting to consider also the over-strong case, $\chi \gg
1$. In this case, the nonlinear frequency is larger than the linear
frequency, so two interacting Alfv\'{e}n wave packets can undergo
multiple cascades to smaller scales in a single linear crossing
time. Therefore, the Alfv\'{e}n wave packets are expected to be rapidly
cascaded in the parallel direction until they restore the condition of
critical balance, $\chi \simeq 1$. Note that fluctuations with
$k_\parallel
\simeq 0$, and therefore with $\omega \simeq 0$, are naturally
generated by three-wave interactions of the \Alfvenic turbulence
\cite{Montgomery:1981,Goldreich:1995}. This 
regime of over-strong turbulence is dominated by uncorrelated
fluctuations, because fluctuations of scale $k_{\perp}$ are
decorrelated over parallel distances greater then $k_{\parallel}\chi$. Thus, the energy in this over-strong region 
of wavenumber space is expected to be roughly constant, as observed in
simulations of MHD turbulence \cite{Maron:2001}.

The assumptions of critical balance and constant energy cascade rate lead to a predicted wavenumber anisotropy scaling of
\begin{equation}
k_\parallel \propto
\begin{cases}
 k_{\perp}^{2/3},&{k}_\perp \rho_i\ll 1\\
 k_{\perp}^{1/3}, &{k}_\perp \rho_i\gg 1,
\end{cases}
\end{equation}
which can be combined into a single equation of the form
\begin{equation}\label{eq:kparkperp}
{k}_\parallel \rho_i= (k_{i}\rho_i)^{1/3} \frac{({k}_\perp\rho_i)^{2/3} + ({k}_\perp\rho_i)^{\xi + 2}}{1 + ({k}_\perp\rho_i)^{2}},
\end{equation}
where $\xi = 1/3$ gives the standard dissipation range scaling derived
from fluid theories assuming critical balance holds in the dissipation
range \cite{Galtier:2006,Howes:2008b,Schekochihin:2009}. The form of Equation~\eqref{eq:kparkperp} has been chosen so that the asymptotic limits are continuously connected and it is well-behaved in its domain. Note that substituting 
Equation~\eqref{eq:kparkperp} into Equation~\eqref{eq:disprel} yields the 
linear Alfv\'{e}n/KAW frequency $\omega$ in terms of only $k_\perp$.

\begin{figure}[h]
\begin{center}	
		\includegraphics[width=\linewidth]{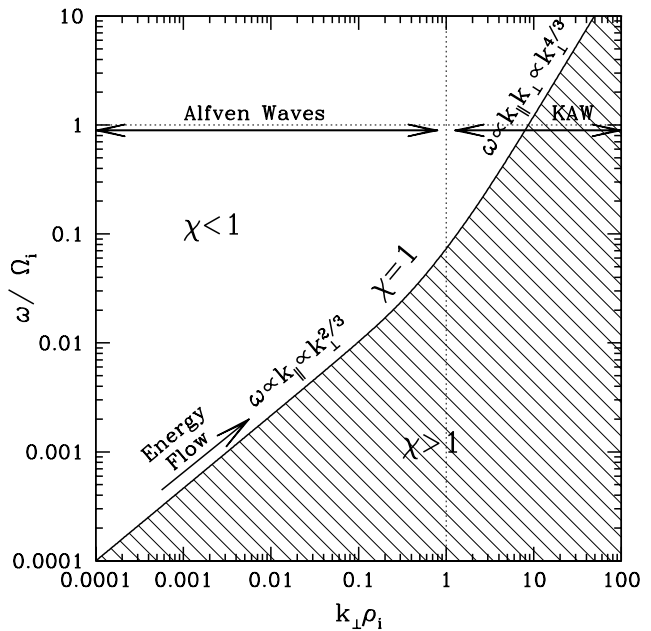}
\end{center}
      \caption{Schematic diagram depicting the population of energy in the $k_{\perp}$-$\omega$ plane assuming critical balance holds. The solid lines correspond to the critical balance boundary. The isotropic energy injection scale is ${k}_{i} \rho_i = 10^{-4}$.}\label{fig:crit_bal}
\end{figure}

A schematic diagram depicting the expected population of  energy in
critically balanced \Alfvenic turbulence in terms of $\omega$ and $k_{\perp}$ is
presented in Figure \ref{fig:crit_bal}. The dispersion relation
defines the critical balance boundary in the $k_{\perp}$-$\omega$ plane
and corresponds to the solid line in Figure
\ref{fig:crit_bal}. Critical balance suggests the turbulent energy
will fill the region below the critical balance boundary (shaded), as seen in
inertial range MHD simulations of turbulence, e.g.,
\cite{Cho:2000,Maron:2001,Cho:2002,Cho:2003,Oughton:2004}, and in
dissipation range electron MHD simulations, e.g.,
\cite{Cho:2004,Cho:2009}.

To obtain a more realistic prediction of critical balance, we construct a
physical model of the expected energy distribution. Previous models of
critical balance, e.g., \cite{Goldreich:1995,Maron:2001}, assume a
form for the energy distribution in $k_\perp$-$k_{\parallel}$ space of
$E(k_{\parallel},k_{\perp}) \sim k_{\perp}^{-10/3} f(k_{\parallel}
L^{1/3}/k_{\perp}^{2/3})$, where $f(|u|) \simeq 1$ for $u \le 1$ and
is negligibly small for $u \gg 1$.

We assume a similar functional form in $k_\perp$-$\omega$ space,
\begin{eqnarray}\label{eq:modelenergy}
\lefteqn{E(\omega,k_{\perp})=} & &  \nonumber \\
& &E_{0} \frac{({k}_\perp\rho_i)^{-\gamma - 2/3} + ({k}_\perp\rho_i)^{-\epsilon + 0.8}}{1 + ({k}_\perp\rho_i)^{2}} \frac{1}{1+\left(\hat{\omega}/\hat{\omega}_{cb}\right)^{8}},
\end{eqnarray}
where $E_{0}$ is an overall energy normalization, $\gamma$ and
$\epsilon$ are the desired inertial and dissipation range energy
scalings, and $\hat{\omega}_{cb} = k_\parallel^2 a_0 \bar{\omega} /
\sqrt{\beta_i}$. The functional form for the $k_\perp$ portion of
Equation~\eqref{eq:modelenergy} is chosen empirically so that the
frequency integrated energy reduces to a one-dimensional perpendicular
energy spectrum that scales as $E_{B_{\perp}}(k_{\perp}) = \int
E(\omega,k_{\perp})\,d\omega\ \propto k_{\perp}^{-\gamma}$ in the
inertial range and $k_\perp^{-\epsilon}$ in the dissipation range. We
take $\gamma = 5/3$ and $\epsilon = 2.8$, consistent with the
dissipation range solar wind \cite{Alexandrova:2009} and large scale kinetic
numerical simulations \cite{Howes:2011b}. The functional form of the
frequency has been chosen to have a flat energy spectrum up to $\sim
\omega_{cb}$ followed by an $8$-th order roll-off at higher
frequencies.

\begin{figure*}[t]
\begin{center}	
	\subfloat[]{\label{fig:cbtheory}\includegraphics[width=0.5\textwidth]{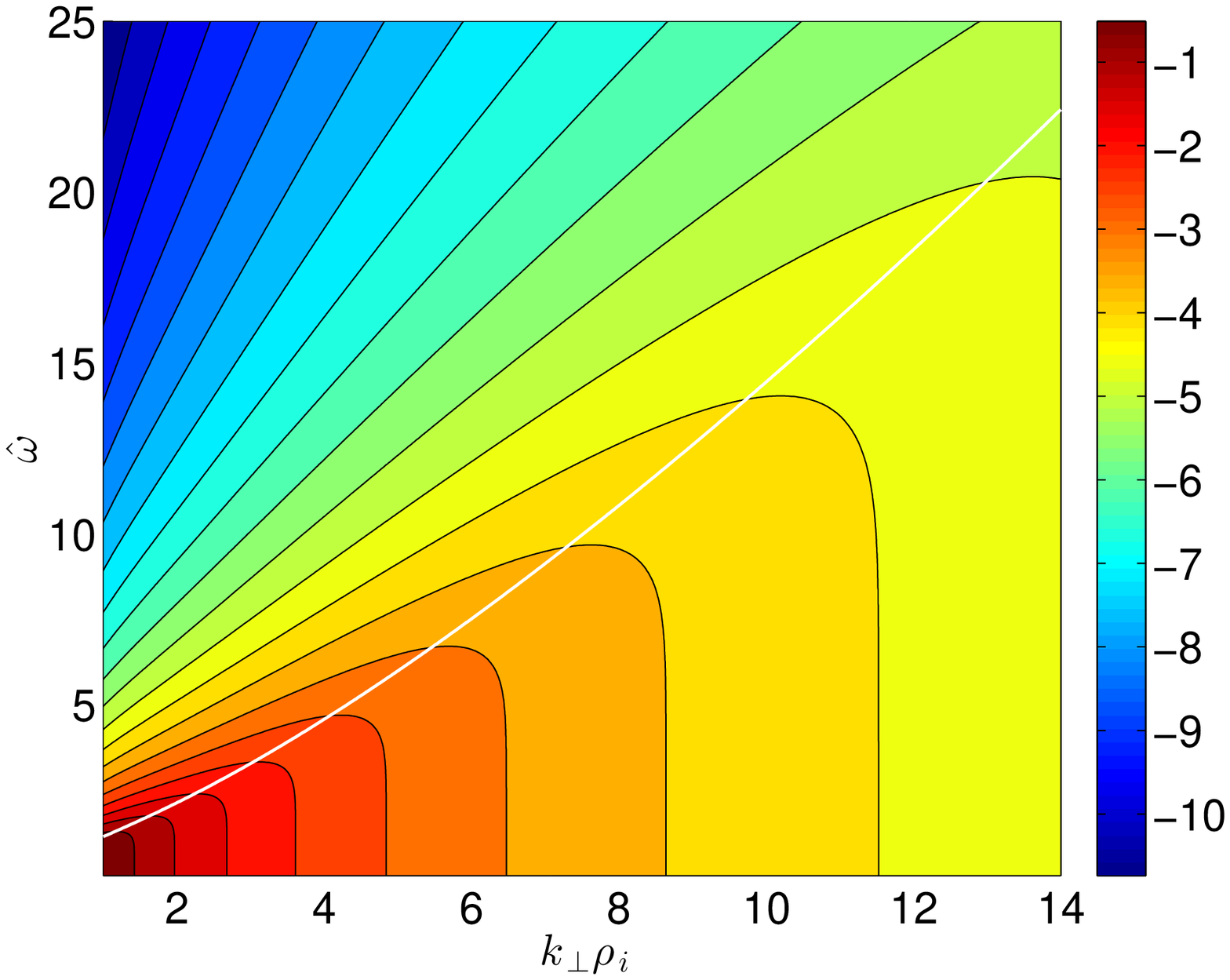}}                
 	\subfloat[]{\label{fig:cbtheoryL}\includegraphics[width=0.5\textwidth]{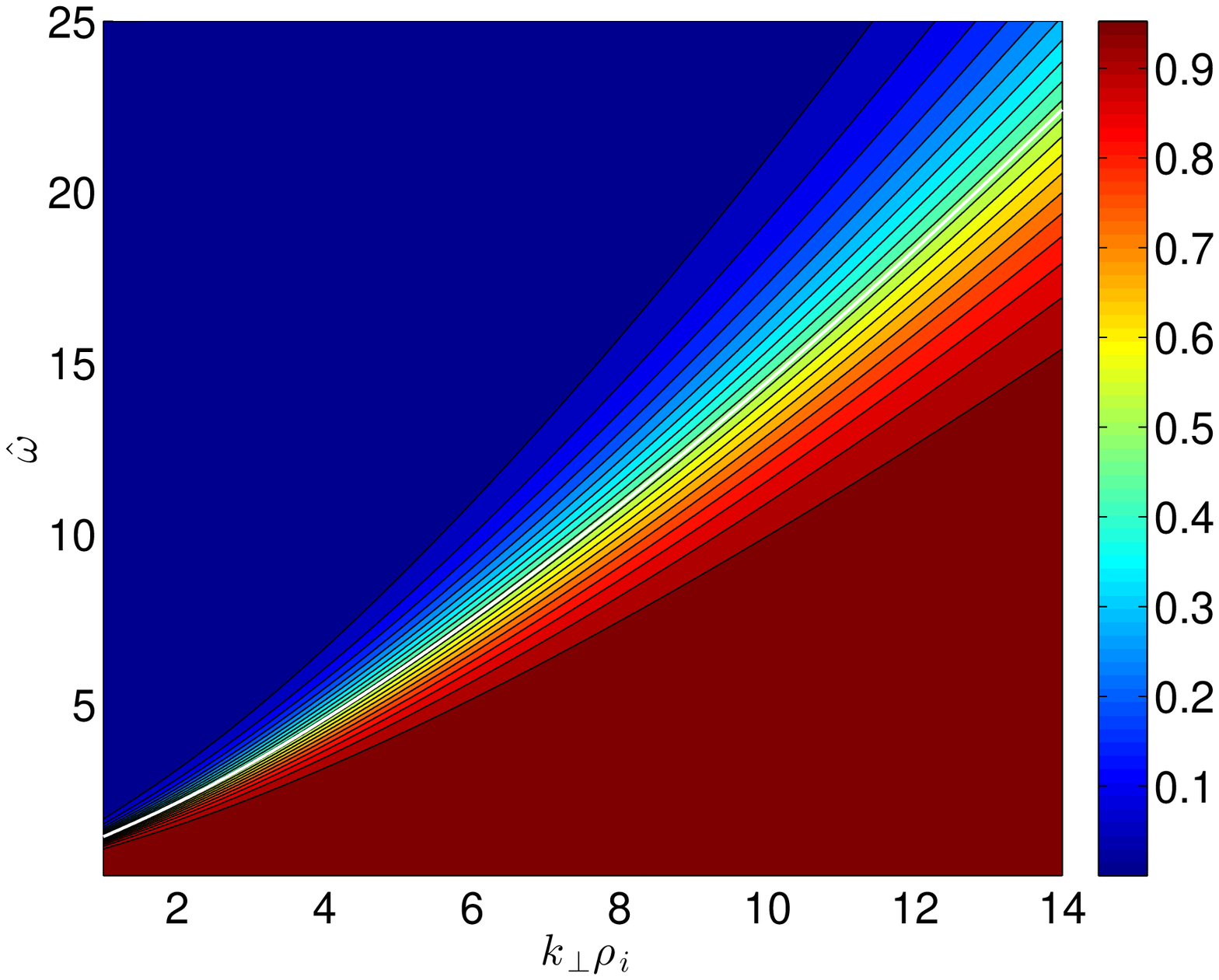}}
\end{center}

  \caption{(Color online) (a) $\log{\left[E_{B_\perp}(\omega,k_\perp)\right]}$ and (b) $E_{B_\perp}(\omega,k_\perp) / E_{B_{\perp}}(0,k_\perp)$, both computed based upon theoretical expectations. The white curve in both panels corresponds to the linear dispersion relation of kinetic Alfv\'{e}n waves with $\xi = 1/3$.}
\end{figure*}

The logarithm of Equation ~\eqref{eq:modelenergy} is plotted in Figure
\ref{fig:cbtheory}. Over plotted in Figure \ref{fig:cbtheory} is the
linear frequency, which corresponds to the critical balance
boundary. To make the boundary more clear, normalizing the energy at
each $k_{\perp}$ by the zero frequency energy,
$E(\omega,k_{\perp})/E(0,k_{\perp})$, yields Figure
\ref{fig:cbtheoryL}.

\begin{figure}[h]
\begin{center}	
		\includegraphics[width=\linewidth]{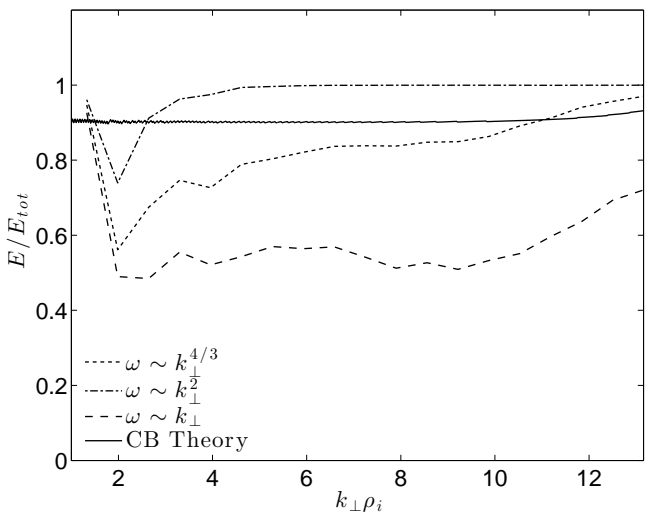}
\end{center}
      \caption{The fraction of energy beneath the critical balance boundary at each $k_{\perp}$. The solid line is based upon theoretical considerations of critical balance, while the other three curves are calculated from the AstroGK simulation with different values of $\xi$.}\label{fig:cbtheoryfraction}
\end{figure}

Since we have now constructed a more realistic numerical model of the
critical balance energy distribution, given by
Equation~\eqref{eq:modelenergy}, we can examine the fraction of the
turbulent energy falling below the critical balance boundary (in
frequency) relative to the total energy at each $k_{\perp}$,
\begin{equation}
\frac{E}{E_{tot}} = \frac{\int_{\omega<\omega_{cb}} E(\omega,k_{\perp})\,d\omega\ }{\int E(\omega,k_{\perp})\,d\omega\ }.
\end{equation}
This fraction is plotted as the solid line in Figure \ref{fig:cbtheoryfraction}. Up to the point where finite box size limitations become important, $\simeq 90\%$ of the energy lies below the critical balance boundary.

\section{Critical Balance Simulation}\label{sec:cbs}
We may now compare the energy distribution in $k_{\perp}$-$\omega$
space from the theoretical model for critical balance, given by
Equation~\eqref{eq:modelenergy}, with that determined through
frequency analysis of our kinetic \Alfven wave turbulence simulation
using AstroGK.

\begin{figure*}[t]
\begin{center}	
	\subfloat[]{\label{fig:contour}\includegraphics[width=0.5\textwidth]{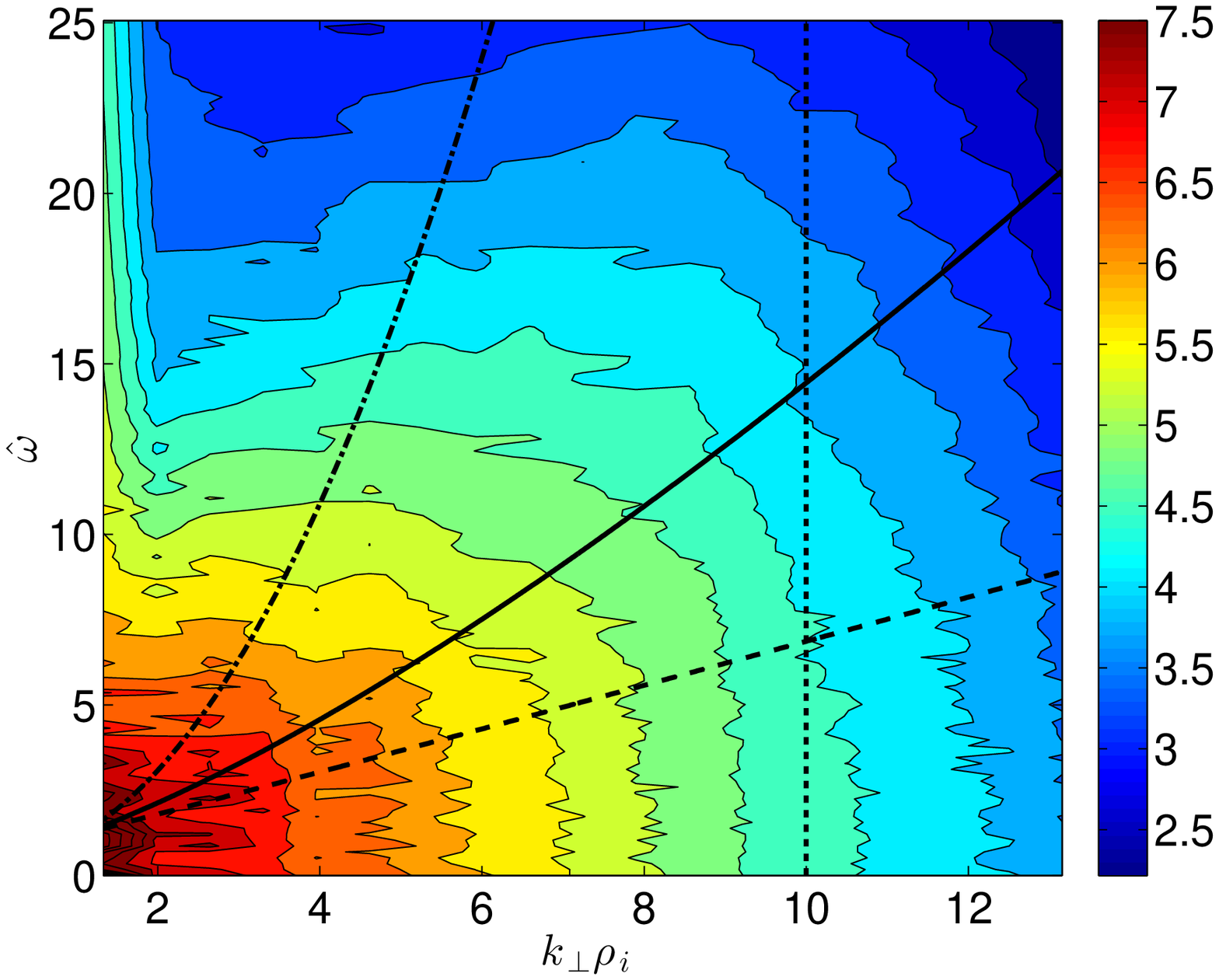}}                
 	\subfloat[]{\label{fig:contourL}\includegraphics[width=0.5\textwidth]{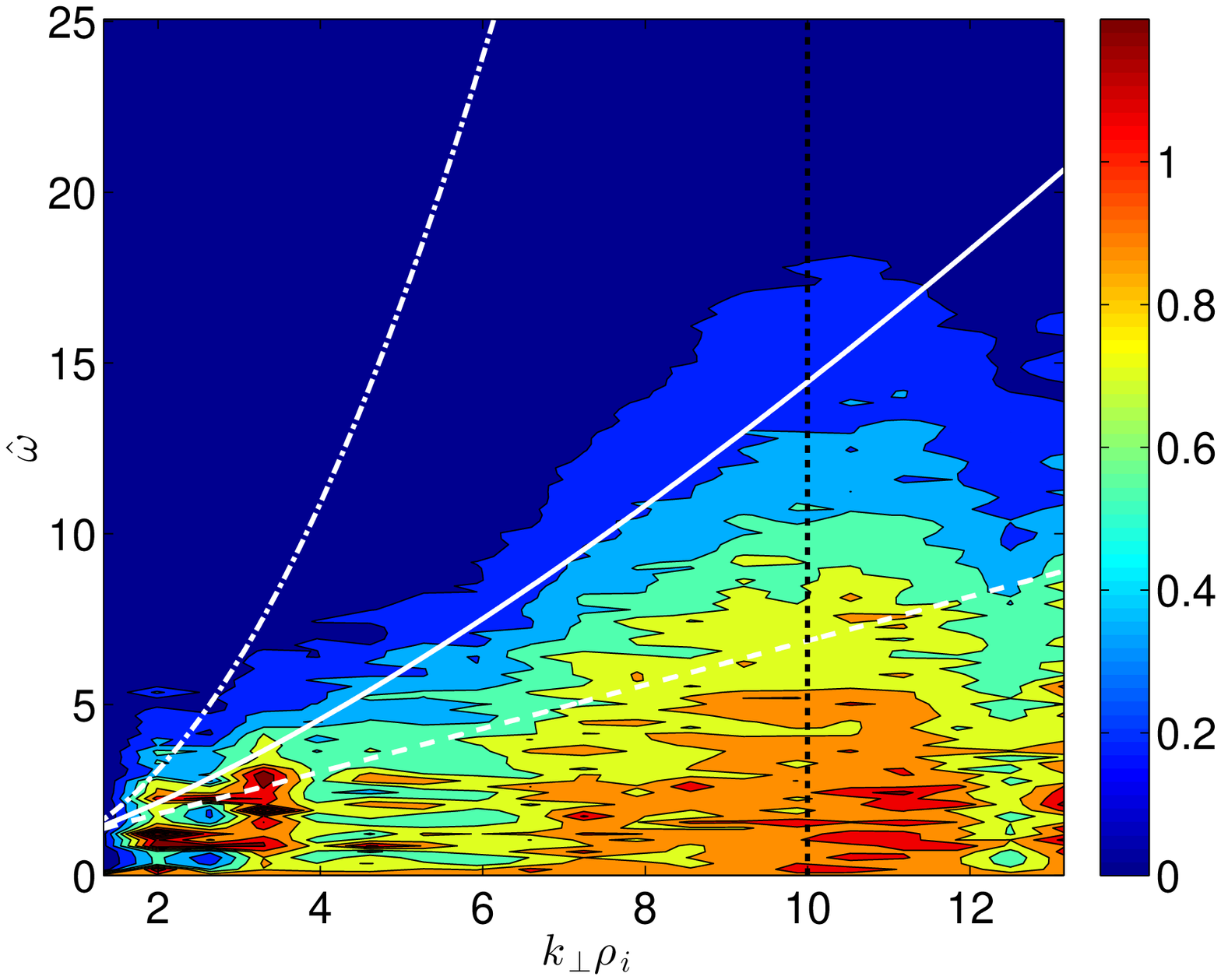}}
\end{center}

  \caption{(Color online) (a) $\log{\left[E_{B_\perp}(\omega,k_\perp)\right]}$ and (b) $E_{B_\perp}(\omega,k_\perp) / E_{B_{\perp}}(0,k_\perp)$ from the AstroGK simulation. The curves correspond to critical balance with $\xi = 0, 1/3,$ and $1$ in ascending order.}
\end{figure*}

In Figure \ref{fig:contour} is plotted the logarithm of the
perpendicular magnetic energy. The figure was constructed by
integrating the 3D AstroGK data in $k_z$ and annular sections in the
$k_x$-$k_y$ plane. The over-plotted curves (black) correspond to
critical balance with $\xi$ in Equation ~\eqref{eq:kparkperp} set to
$0$ (dashed), $1/3$ (solid), and $1$ (dot-dashed), where $\xi = 1/3$
is the conventional prediction for critically balanced kinetic \Alfven
wave turbulence in the dissipation range.

Below the $\xi = 1/3$ (solid black) critical balance boundary in
Figure \ref{fig:contour}, we find excellent qualitative agreement
with the theoretical prediction represented in Figure
\ref{fig:cbtheory}. Although the agreement is slightly poorer above
this boundary, the fraction of the energy in this region is very
small. The energy fraction falling below each of the curves in Figure
\ref{fig:contour} is plotted in Figure \ref{fig:cbtheoryfraction}. The
upward trend beginning at $k_\perp \rho_i \simeq 8$ is due to
hypercollisionality and finite box size limitations, and the poor
agreement at small $k_\perp$ is due to the effect of driving. Away
from these limiting values, the difference between the theory and the
simulation for $\xi = 1/3$ is within approximately $10\%$.

In Figure \ref{fig:contourL} is plotted the perpendicular magnetic
energy at each $k_\perp$ normalized to the zero frequency energy,
$E_{B_\perp}(\omega,k_\perp) / E_{B_{\perp}}(0,k_\perp)$. The figure
was constructed in the same manner as Figure \ref{fig:contour}, and
the curves (white) correspond to the same values of $\xi$. As
demonstrated with the theoretical model, plotting the energy
distribution in $\omega$-$k_{\perp}$ space normalized in this fashion
highlights the critical balance boundary, which closely follows the
standard dissipation range prediction of $\xi = 1/3$ (solid
white). Again, the qualitative agreement with the prediction of
critical balance is excellent. The non-smooth distribution of energy
below the critical balance line in Figure
\ref{fig:contourL} is primarily due to the discrete nature of this 
moderate resolution simulation.

\begin{figure}[h]
\begin{center}	
		\includegraphics[width=\linewidth]{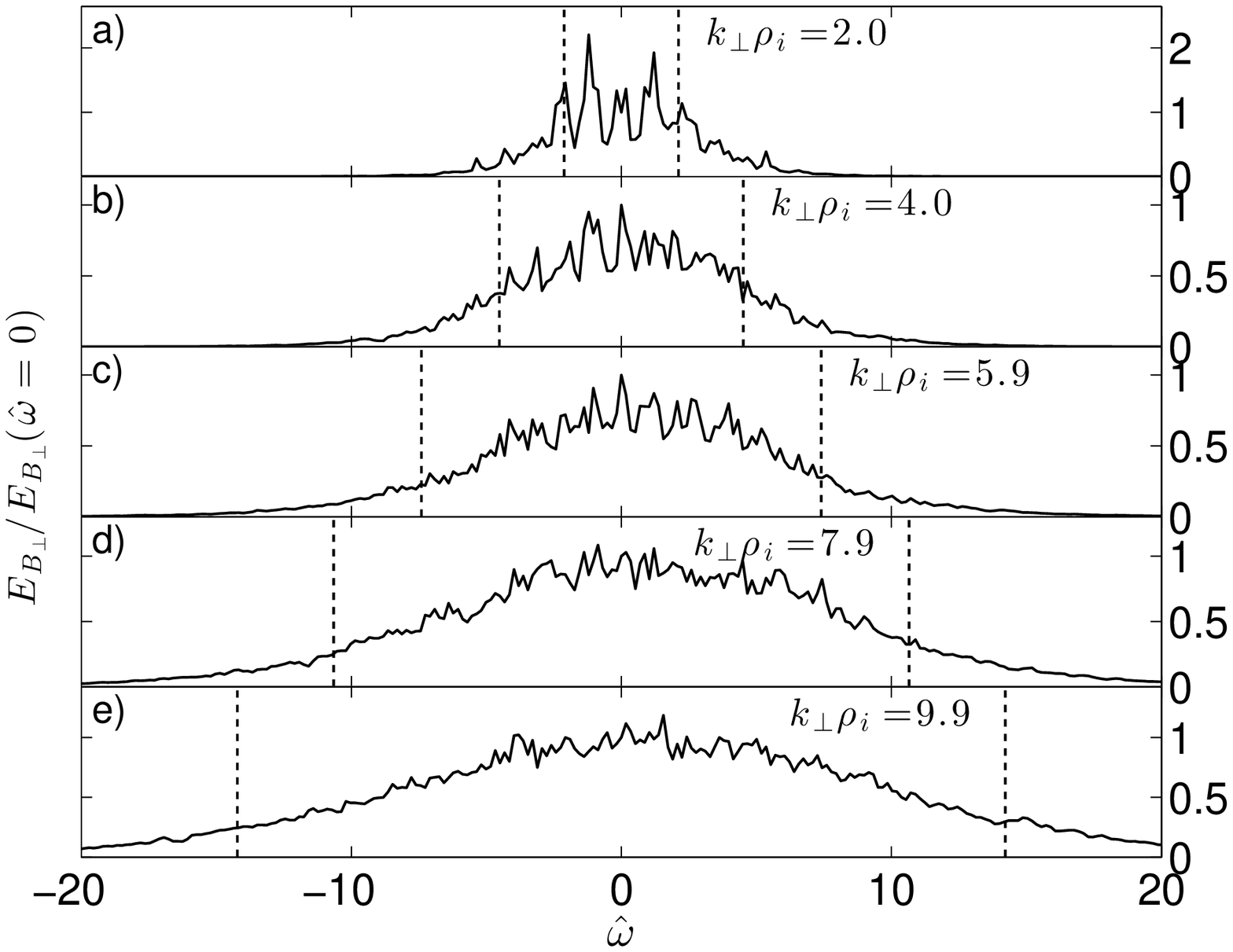}
\end{center}
      \caption{Perpendicular magnetic energy normalized to the zero
      frequency energy at each $k_\perp$ versus frequency at fixed
      $k_\perp$. The vertical lines represent the $\xi = 1/3$ critical
      balance boundary. Panels a) - e) step through $k_\perp$ bins.
      }\label{fig:cutoff}
\end{figure}

Another method of visualizing the perpendicular magnetic energy
distribution is to perform cuts along $\hat{\omega}$ at fixed
$k_\perp$, as presented in Figure \ref{fig:cutoff}. Since the energy is
plotted linearly, the area under each curve corresponds to the
turbulent energy. Each panel of the figure represents a different
$k_\perp$ slice, and the vertical dashed lines indicate the $\xi =
1/3$ critical balance boundary. Panel a is influenced by the driving
at $k_\perp \rho_i=1$, but it still displays similar qualitative
behaviour to cuts at higher $k_\perp \rho_i$. The general trend can be
seen to be an approximately flat energy distribution up to a frequency
somewhat less than the critical balance boundary, followed by a steep
roll-off. The majority of the energy in all cases is contained within
the critical balance boundary. This plot makes clear an important
characteristic of the turbulence: no energy significantly in excess of
that predicted by the critical balance model is seen either at
$\hat{\omega} = 0$ or at frequencies above critical balance.

\section{Eulerian Frequency Analysis}\label{sec:eul}

\begin{figure}[h]
		\includegraphics[width=\linewidth]{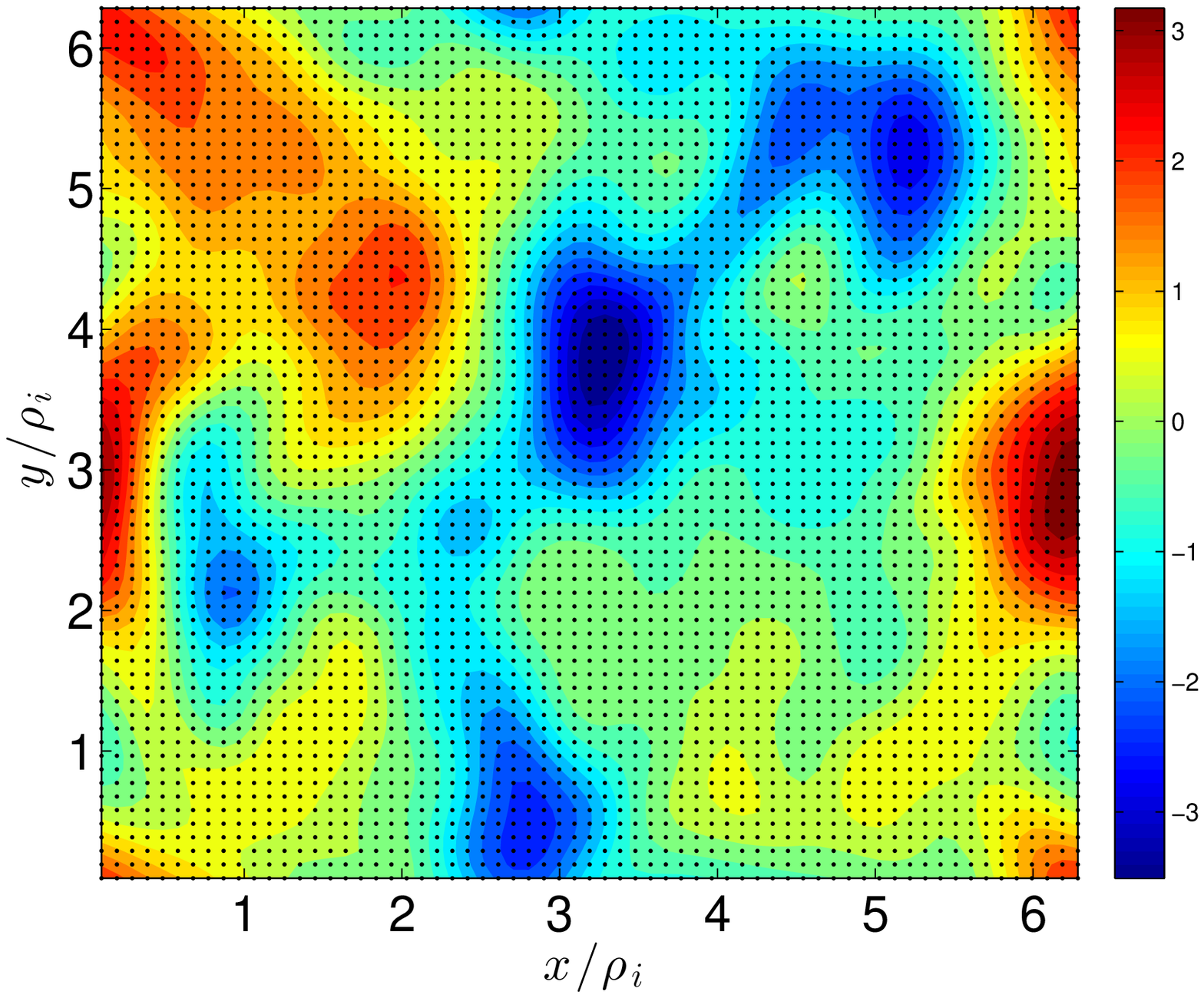}
      \caption{(Color online) $E_{B_x}$ at a single $z$ plane in the middle
      of the simulation domain. Overlaid are the positions of the
      ``probes" used to record temporal
      fluctuations.}\label{fig:eulbx}
\end{figure}

Eulerian frequency spectra are constructed by placing an array of
``probes" across a single spatial $x$-$y$ plane in the middle of the
simulation domain. We use an array of $64 \times 64$ probes across
the plane to record a time series of the fluctuating magnetic field
components. A schematic of the probe distribution can be seen in
Figure \ref{fig:eulbx}.

\begin{figure}[h]
\begin{center}	
		\includegraphics[width=\linewidth]{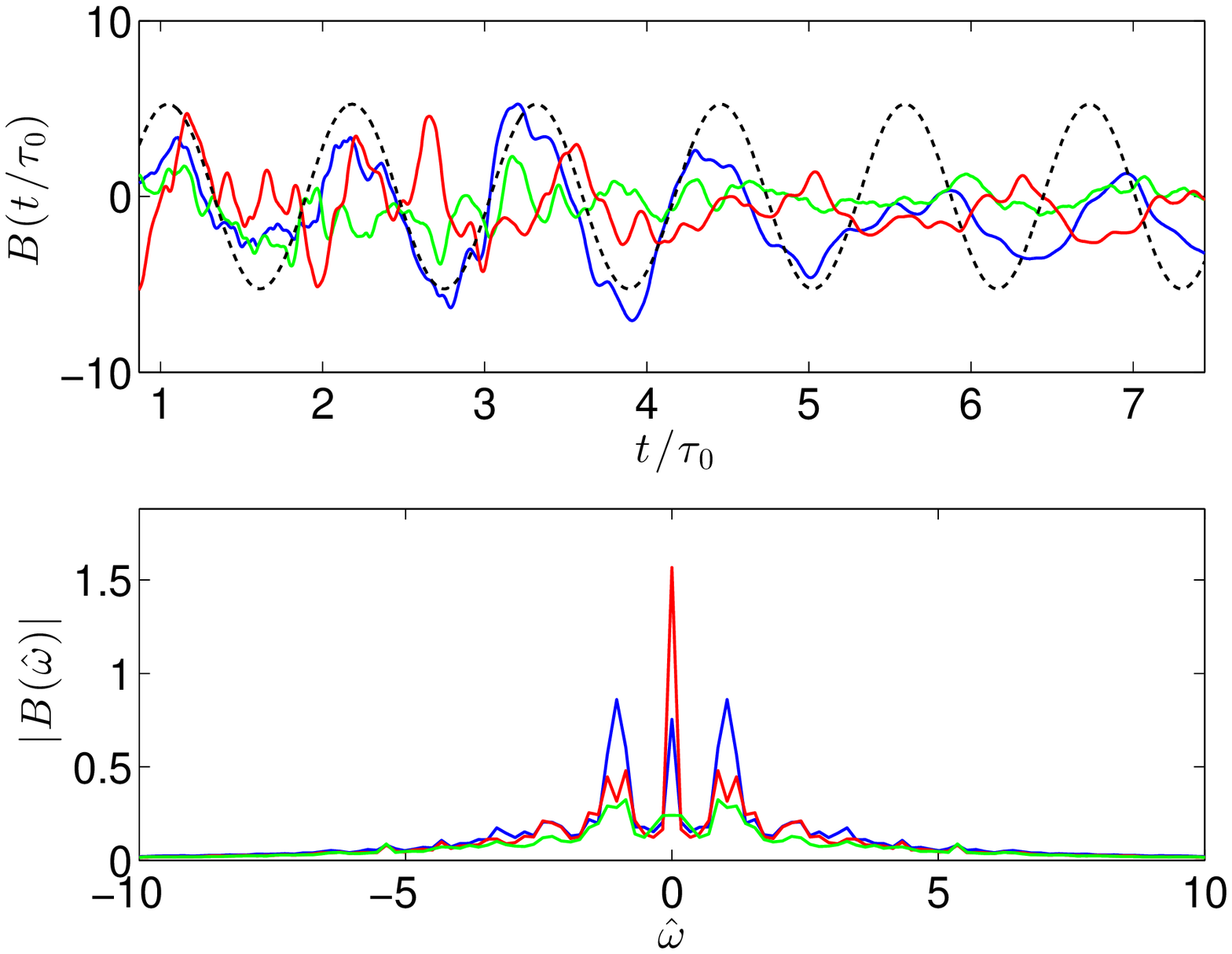}
\end{center}
      \caption{(Color online) In the upper panel is presented the temporal
      evolution at a single probe location of the three magnetic field
      components: $B_x$ (blue), $B_y$ (red), and $B_z$ (green). The
      dotted black line is $\cos{(t/\tau_{0} + 0.5)}$. In the lower panel
      are presented the Fourier responses of the three magnetic field
      components averaged over the full probe array. }\label{fig:eult}
\end{figure}

The evolution in time of the three magnetic field components at a
single probe location over the same time interval used for the
preceding critical balance study are illustrated in the upper panel of
Figure \ref{fig:eult}. For illustrative purposes, over-plotted in the figure
is $\cos{(t/\tau_{0} + 0.5)}$. Clearly, the $B_x$ (blue) component is
dominated by $\hat{\omega} \simeq 1$ fluctuations, which corresponds
to the driving frequency and the lowest linear mode of the system.

In the lower panel of Figure \ref{fig:eult} are plotted the Fourier
response of the three magnetic field components averaged over the full
probe array. All three magnetic field components are dominated by
spectral peaks at $\hat{\omega} \simeq 0$ and $\pm 1$, with a clear
band-gap between the values. As noted above, $\hat{\omega} \simeq
\pm 1$ corresponds to the driving frequency and the lowest linear mode of
the system. As such, this spectral component should contain the most
power due to the driving, responsible for generating the forward
cascade of energy to higher frequencies. We conjecture that the peak
at $\hat{\omega} = 0$ exhibits significant energy because this mode is
responsible for nonlinear scattering in three-wave interactions of
turbulence and is self-consistently generated via the nonlinear
interaction \cite{Montgomery:1981,Sridhar:1994,Galtier:2003}.

\begin{figure}[h]
\begin{center}	
		\includegraphics[width=\linewidth]{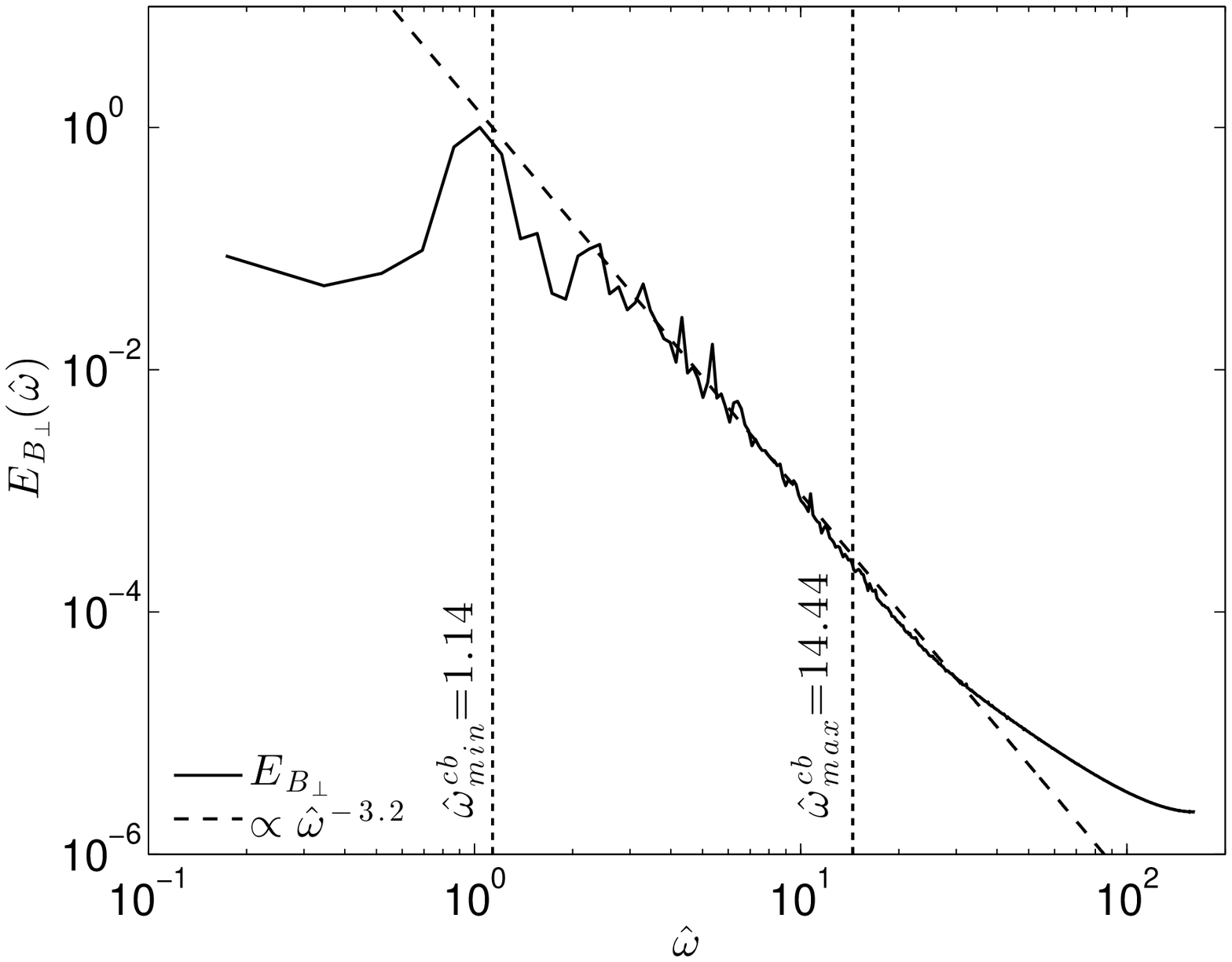}
\end{center}
      \caption{The perpendicular magnetic energy spectrum as a
      function of frequency averaged over the full probe array. The
      dashed line corresponds to a spectral index of $-3.2$. The
      vertical dotted lines correspond to the expected minimum and
      maximum frequencies from linear theory.}\label{fig:euls}
\end{figure}

Figure \ref{fig:euls} presents the Eulerian frequency spectrum of the
perpendicular magnetic energy averaged over all probes.  The vertical
dotted lines indicate the minimum and maximum linear frequencies for
the minimum and maximum resolved length scales in the simulation
domain, as given by Equation~\eqref{eq:disprel}. The frequency range
around $\hat{\omega} \simeq 1$ is dominated by the antenna driving, so
we fit from $\hat{\omega} = 2$ to $14.4$ to obtain a spectral index of
$\simeq -3.2$ for the Eulerian frequency spectrum.

Excluding the the driving dominated portion of the spectrum around
$\hat{\omega} \simeq 1$, the energy in the low frequency range below
$\hat{\omega}^{cb}_{min}$ is approximately constant, consistent with
the predictions of critical balance outlined in \S \ref{sec:cbt}. It
is important to note that very little energy resides in these low
frequency modes: the total integrated energy in the turbulence is the
area under a linear plot of the frequency spectrum, so this
low-frequency range corresponds to very little integrated area but is
exaggerated in the logarithmically plotted Figure~\ref{fig:euls}.

\section{Discussion}\label{sec:disc}

The model for critical balance constructed in \S \ref{sec:cbt} is
based upon the theoretical expectations for strong turbulence, the
foundation for which was established by \citet{Goldreich:1995}: our
model quantifies the qualitative arguments given therein and extends
the notion of critical balance into the dissipation range, permitting
arbitrary perpendicular wavevector scaling to agree with the results of
large-scale kinetic numerical simulations \cite{Howes:2011b} and solar
wind observations \cite{Alexandrova:2009} of the dissipation range.

The theoretical model of critical balance is compared to our numerical
simulations in \S \ref{sec:cbs}. Considering that the concept of
critical balance is based upon order of magnitude scaling arguments,
the agreement between theory and simulation presented in Figure
\ref{fig:contourL} is striking. It is important to note that the
nature of the energy input into our simulation domain generates
perpendicular magnetic energy fluctuations that resemble linear
Alfv\'{e}n/KAWs. The turbulent fluctuations, however, are only driven
at the smallest resolved wave numbers of our simulation domain, with
the frequency range of the driving energy centered at $\hat{\omega} =
\hat{\omega}_0$. Therefore, all of the energy at larger wave numbers and
frequencies away from $\hat{\omega}_0$ results from self-consistent,
nonlinear turbulent interactions.

The spectral anisotropy in fluid simulations is typically determined
via two-point, second-order structure function analyses, e.g.,
\cite{Cho:2000,Maron:2001,Cho:2004,Chen:2011a}. This approach allows
one to identify the wavevector parallel to the \emph{local} mean
magnetic field at a given scale. A standard Fourier analysis
provides only wavevectors parallel and perpendicular to the \emph{global}
mean magnetic field, i.e., $k_z$ and $k_R$, rather than $k_\parallel$
and $k_\perp$
\cite{Cho:2000,Maron:2001}. This effect is a consequence of the
inherent averaging of the Fourier transform. Turbulent fluctuations
relevant to solar wind turbulence in the dissipation range are
characterized by $\delta B_\perp / B_0 \ll 1$ and $k_\parallel /
k_\perp \ll 1$, so the Fourier transformed wavevector components are
given, to lowest order, by $k_R
\simeq k_\perp$ and $k_z \simeq \theta k_\perp$, where $\theta \simeq
\delta B_\perp / B_0$ \cite{Maron:2001}. Thus, a standard Fourier
approach to analyzing turbulent simulations is sensitive only to the
spectrum in $k_\perp$ and the amplitude of the fluctuating field, rather than $k_\parallel$ spectrum along the local mean magnetic
field.

The structure function approach works well in MHD inertial range
turbulence simulations and undamped dissipation range turbulence
simulations, wherein the spectral slope is relatively shallow.  The
perpendicular spectra in kinetic simulations of dissipation range
turbulence \cite{Howes:2011b} and in solar wind observations
\cite{Alexandrova:2009} have spectral indices around  $-2.8$, and the
parallel spectrum is expected to be steeper yet
\cite{Cho:2004,Schekochihin:2009,Chen:2010a}. Two-point, second-order
structure functions cannot recover spectral indices steeper than $-3$
\cite{Farge:2006},  and thus structure functions cannot be used to 
analyze the parallel spectrum of our dissipation range simulations.

The analysis performed herein obviates the limitations of structure
function and spatial Fourier analyses by using the frequency $\omega$
as a proxy for $k_\parallel$. This approach is motivated by the fact
that the linear frequency for the \Alfven and kinetic \Alfven
wave is linearly proportional to parallel wavenumber, $\omega \propto
k_\parallel$.  This relation can be seen clearly by rewriting
Equation~\eqref{eq:disprel} in the form $\omega(k_\perp,k_\parallel)=
\overline{\omega}(k_\perp) k_\parallel v_A$. The observation of a
critical balance scaling in the turbulent energy distribution in
$k_{\perp}$-$\omega$ space for our kinetic simulation of dissipation
range turbulence implies that the spectral anisotropy observed in MHD
inertial range simulations
\cite{Cho:2000,Maron:2001,Cho:2002,Cho:2003} and electron MHD 
dissipation range simulations \cite{Cho:2004,Cho:2009} persists even
in a turbulent kinetic plasma. Since the conventional critical balance
boundary given by $\xi = 1/3$ is a good fit to the turbulent energy
distribution from our simulation, we conclude that the turbulent
spectrum of kinetic \Alfven waves is well described by an anisotropic,
critically balanced energy distribution in wavevector space given by $0
<\xi \le 1/3$, even for a fully nonlinear and collisionlessly damped
kinetic simulation.

The results of \S \ref{sec:cbs}
 and \ref{sec:eul} 
demonstrate that the theory of critical balance extends into the
dissipation range of turbulence. Since critical balance is
fundamentally a balancing of the linear and nonlinear processes in
plasma turbulence, our results suggest that linear wave theory plays
an important role, even in a strongly nonlinear turbulent plasma.

\citet{Dmitruk:2009} (DM) performed a similar Eulerian frequency
analysis of driven MHD turbulence, but their conclusions about the
nature of the turbulence differ dramatically from our findings.  In
particular, they claim to find an excess accumulation of energy
extending from their lowest linear mode to zero frequency and a very
steep spectrum above their lowest linear mode. They conclude that
linear waves play little role in MHD plasma turbulence. We believe
the conclusions of DM are significantly biased by the set-up of their
numerical simulations, for  a number of reasons detailed below.

First, DM drive their simulations isotropically (with $k_\perp\sim
k_\parallel$ for the driven modes) with a fixed amplitude, yielding a
turbulent fluctuation amplitude $\delta B \sim 1$, for a range of
values of the mean field strength $B_0=0,1,2,8,16$. The nonlinearity
parameter for Alfv\'{e}nic turbulence may be expressed as $\chi =
k_\perp \delta B / k_\parallel B_0$, so the \emph{only} case that
yields strong MHD turbulence with $\chi \sim 1$ is the $B_0=1$ case;
all other cases correspond to simulations of weak turbulence, with
increasingly small nonlinearity parameters, $\chi =1/2,1/8,1/16$.
That the turbulence is indeed weak is supported by the very steep
spectral indices of their frequency spectra.

The nonlinear frequency in the weak turbulence
regime\cite{Maron:2001,Howes:2011c} is given by $\omega_{nl} \simeq \chi^2
k_\parallel v_A$ , which suggests that one may expect to see a signature in the
frequency spectrum corresponding to this very low nonlinear frequency,
as indeed observed by DM. In addition, DM report that their
simulations require ``tens of nonlinear times'' to reach a saturated
steady state, although they do not define what they mean by a
``nonlinear time.''  If nonlinear time is taken to mean the \Alfven crossing time, their result is consistent with the expectation that it
will take approximately one full nonlinear timescale, $\tau_{nl}
\simeq \tau_A / \chi^2$, or many \Alfven crossing times,
$\tau_A=2\pi/k_\parallel v_A$, to reach the steady state of the weak
turbulent cascade. Note that other simulations of driven, strong MHD
turbulence report the establishment of a steady state within one to a few
\Alfven times, as expected for $\chi \sim 1$ \cite{Biskamp:1999,Cho:2000,Cho:2004,Cho:2009,Perez:2010a}.
Therefore, we speculate that the excess energy at low frequency
observed in the weak turbulence simulations of DM may be attributed either to the response of the plasma at the low nonlinear frequency or
to the inclusion, in the interval used for the frequency analysis, of
the long transient evolution of the turbulence as it approaches a
steady state.

Second, in weak turbulence, the resonant matching conditions for the
nonlinear term in wavevector and frequency
\cite{Shebalin:1983,Sridhar:1994} for the dominant three wave interactions
\cite{Montgomery:1995,Ng:1996,Goldreich:1997,Galtier:2000,Galtier:2003}
imply a crucial role for modes with $k_\parallel=0$ in the nonlinear
transfer of energy.  According to the linear dispersion relation for
\Alfven waves, $\omega =\pm k_\parallel v_A$,
a fluctuation with $k_\parallel=0$ also has $\omega=0$. If such modes
are generated by the nonlinear interactions in the turbulence, this
could be another possible cause for an excess of energy in their frequency
spectra at $\omega\simeq 0$.

Third, DM report that the frequency spectrum of their driving has the
form $P(\omega)\sim 1/(\omega^2+\omega_c^2)$, with $\omega_c=3$ in
their dimensionless units. The antenna power therefore scales
$P\propto \omega^0$ for $\omega \ll \omega_c$ and $P\propto
\omega^{-2}$ for $\omega \gg \omega_c$. However, DM do not present
a plot of the frequency spectrum of their driving, making it difficult
to assess fairly the impact of the driving on the frequency spectra of
their turbulence simulations. Since the parallel energy spectrum of
strong MHD turbulence \cite{Howes:2011c} is predicted to scale as
$E(k_\parallel) \propto k_\parallel^{-2}$, the linear relationship
between the parallel wavenumber and frequency implies a frequency
spectrum $E(\omega) \propto \omega^{-2}$. In weak turbulence, the frequency spectrum would be steeper yet. Therefore, it is questionable whether one could observe
even the strong turbulent frequency spectrum in the presence of their
driving.

One may attempt to judge the impact of the driving by examining
Figure~8 in DM, a comparison of the frequency spectra from a driven
and an undriven simulation.  It is important to note that both are
weak turbulence simulations, so the contribution to the spectra due to
the nonlinear turbulent fluctuations should be similar. The driven
simulation shows a significantly larger signal over the frequency
range $0.4 \le \omega
\le 10$, suggesting that the driving has a significant, if not
dominant, effect on the frequency spectrum over this range. DM
attribute the broadband nature of the frequency spectra in their suite
of simulations, presented in their Figure~2, to the inherently
nonlinear nature of MHD turbulence. We suggest that a more careful
evaluation of the impact of their driving on the frequency spectrum is
required to establish the validity of their conclusion.

We believe the arguments outlined above raise serious questions about
the conclusions that DM reach regarding the nature of MHD turbulence,
in particular the claim that linear mode properties play little role
in the turbulent evolution. The concept of critical balance implicitly
assumes that linear wave modes do play a role in plasma
turbulence. The numerical evidence presented here for critical balance
in the kinetic \Alfven wave cascade of dissipation range
turbulence therefore indirectly supports the notion that linear wave
modes do indeed play a role in strong plasma turbulence.

\section{Conclusion}\label{sec:conc}

We have developed a theoretical model for critically balanced Alfv\'{e}n/KAW turbulence and compared the results of a fully nonlinear, driven, gyrokinetic simulation to the theoretical prediction. We find excellent qualitative and quantitative agreement with the predictions of critical balance in the dissipation range of KAW turbulence. This result constitutes the first evidence of critical balance to be observed in a damped and dissipative kinetic turbulence simulation.

Having found agreement with critical balance implies the anisotropic cascade of Alfv\'{e}n waves in the inertial range extends into the dissipation range, where the anisotropic scaling of KAW turbulence is observed to be approximately $k_\parallel \propto k_\perp^{\xi}$ with $0 < \xi \le 1/3$. The upper bound of this result agrees with theoretical predictions for the dissipation range scaling based upon fluid descriptions, and the damping present in our kinetic simulation is expected to strengthen the anisotropy \cite{Howes:2011c}.

The results of our Eulerian frequency analysis performed by temporally sampling the magnetic field data across a fixed $x$-$y$ plane provide further evidence of a critically balanced cascade of energy beginning at our driving frequency. Aside from the expected population of energy in the $\omega = 0$ mode, we find no evidence of excess energy either below the lowest or above the highest linear modes resolved in our simulation, in contradiction to \citet{Dmitruk:2009}. 

Although the current analysis does not directly address the importance linear wave theory to plasma turbulence, agreement with critical balance implies linear wave modes play some role in governing turbulence. Forthcoming simulation analyses will explore the importance of linear wave modes in fully developed, strong plasma turbulence in more detail.

\begin{acknowledgements}

J.M.T.  and G.G.H. thank ISSI team 185 ``Dispersive cascade and
dissipation in collisionless space plasma turbulence---observations
and simulations.''  Support was provided by the International Space
Science Institute in Bern, NSF CAREER Award AGS-1054061, and NSF grant
PHY-10033446.

\end{acknowledgements}

\end{document}